%% file: main.tex
\documentclass[conference]{IEEEtran}

\usepackage{cite}
\usepackage{amsmath,amssymb,amsfonts}
\usepackage{graphicx}
\usepackage{textcomp}
\usepackage{xcolor}

\usepackage{balance}
\usepackage{multirow}
\usepackage{hyperref}
\usepackage{dblfloatfix}
\usepackage{algorithm}
\usepackage[noend]{algpseudocode}
\usepackage[normalem]{ulem}

\usepackage[firstpage]{draftwatermark}
\SetWatermarkText{
\hspace*{6.5in}\raisebox{10.25in}{\href{https://www.acm.org/publications/policies/artifact-review-and-badging-current}{\includegraphics[scale=.2]{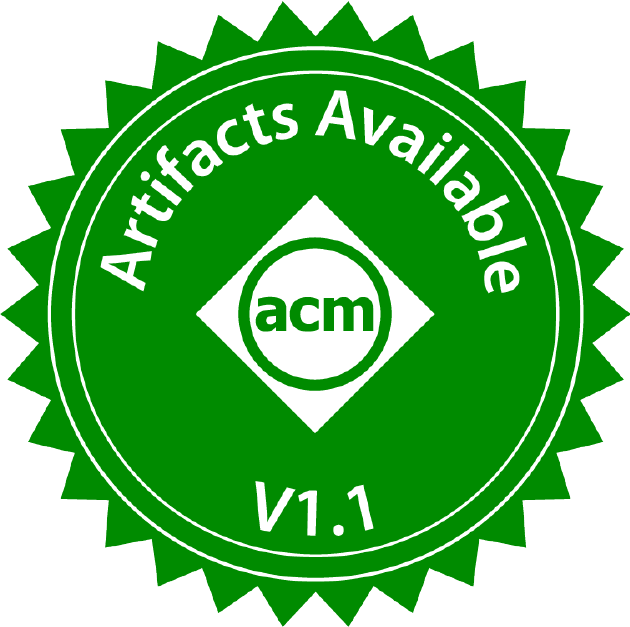}}\href{https://www.acm.org/publications/policies/artifact-review-and-badging-current}{\includegraphics[scale=.2]{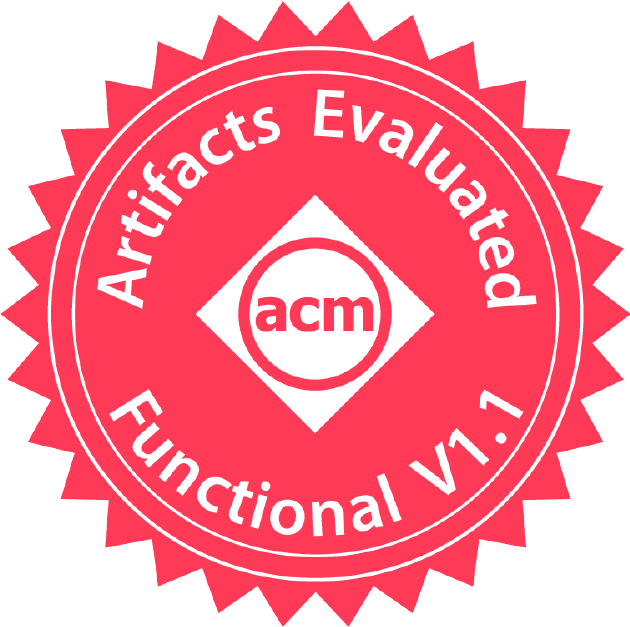}}\href{https://www.acm.org/publications/policies/artifact-review-and-badging-current}{\includegraphics[scale=.2]{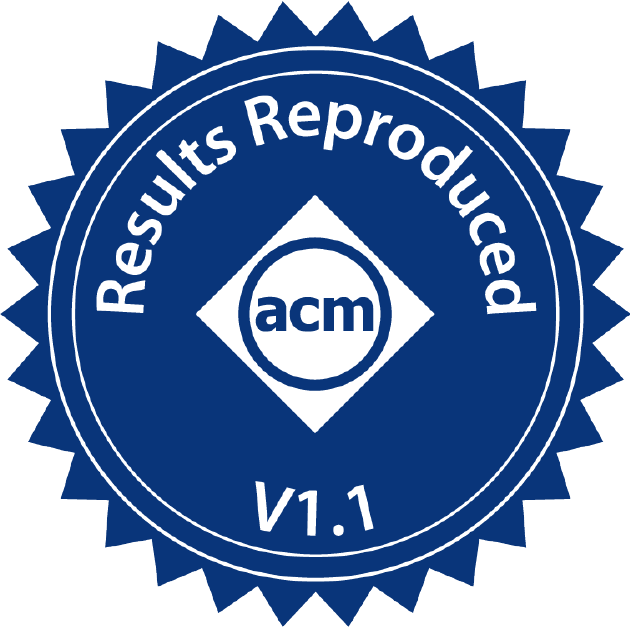}}}}
\SetWatermarkAngle{0}

\def\BibTeX{{\rm B\kern-.05em{\sc i\kern-.025em b}\kern-.08em
    T\kern-.1667em\lower.7ex\hbox{E}\kern-.125emX}}

\begin{document}

\title{Parallelizing Maximal Clique Enumeration on GPUs}

\input{authors}

\maketitle

\input{sec/0-abstract}
\input{sec/1-intro}
\input{sec/2-background}
\input{sec/3-approach}
\input{sec/4-evaluation}
\input{sec/5-related}
\input{sec/6-conculsion}
\input{sec/ack}
\balance
\input{sec/artifact}
\bibliographystyle{IEEEtran}
\bibliography{main}

\end{document}

%% file: authors.tex
\author{
\IEEEauthorblockN{Mohammad Almasri\IEEEauthorrefmark{1}\IEEEauthorrefmark{2}, Yen-Hsiang Chang\IEEEauthorrefmark{1}\IEEEauthorrefmark{2}, Izzat El Hajj\IEEEauthorrefmark{3}, Rakesh Nagi\IEEEauthorrefmark{4}, Jinjun Xiong\IEEEauthorrefmark{5}, Wen-mei Hwu\IEEEauthorrefmark{2}\IEEEauthorrefmark{6}}

\IEEEauthorblockA{\IEEEauthorrefmark{1}Both authors contributed equally to this research.}

\IEEEauthorblockA{\IEEEauthorrefmark{2}ECE,\IEEEauthorrefmark{4}ISE,  University of Illinois at Urbana-Champaign, Urbana, IL, USA}

\IEEEauthorblockA{\IEEEauthorrefmark{3}Department of Computer Science, American University of Beirut, Beirut, Lebanon}

\IEEEauthorblockA{\IEEEauthorrefmark{5}Department of Computer Science and Engineering, University at Buffalo, Buffalo, NY, USA}

\IEEEauthorblockA{\IEEEauthorrefmark{6}Nvidia Corporation, Santa Clara, CA, USA}

\IEEEauthorblockA{almasri3@illinois.edu, yhchang3@illinois.edu, izzat.elhajj@aub.edu.lb, nagi@illinois.edu, }

\IEEEauthorblockA{jinjun@buffalo.edu, w-hwu@illinois.edu}
}

%% file: sec/0-abstract.tex
\begin{abstract}

We present a GPU solution for exact maximal clique enumeration (MCE) that performs a search tree traversal following the Bron-Kerbosch algorithm.
Prior works on parallelizing MCE on GPUs perform a breadth-first traversal of the tree, which has limited scalability because of the explosion in the number of tree nodes at deep levels.
We propose to parallelize MCE on GPUs by performing depth-first traversal of independent subtrees in parallel.
Since MCE suffers from high load imbalance and memory capacity requirements, we propose a worker list for dynamic load balancing, as well as partial induced subgraphs and a compact representation of excluded vertex sets to regulate memory consumption.
Our evaluation shows that our GPU implementation on a single GPU outperforms the state-of-the-art parallel CPU implementation by a geometric mean of 4.9$\times$ (up to 16.7$\times$), and scales efficiently to multiple GPUs.
Our code has been open-sourced to enable further research on accelerating MCE.

\end{abstract}

%% file: sec/1-intro.tex
\section{Introduction} \label{sec:intro}

A clique in a graph is a complete subgraph where every vertex in the subgraph is adjacent to every other vertex.
A maximal clique is a clique that cannot be further expanded by including one more vertex.
Maximal clique enumeration (MCE) aims to find all the maximal cliques in a graph, which has a wide variety of applications in numerous domains such as
    community detection~\cite{cd2, cpm1, cpm2, conductance},
    recommender systems~\cite{rs1, rs2},
    graph compression and partitioning~\cite{gc1, gc2, gc3, scalableGPM},
    prediction of protein functions in protein interaction networks~\cite{clique_ppi1, clique_ppi2, clique_ppi3, clique_ppi4},
    finding gene similarities in gene co-expression networks~\cite{clique_gene1, clique_gene2},
    and identifying price fluctuations in finance networks~\cite{clique_finance}.

One of the most widely used algorithms for solving MCE exactly is the Bron-Kerbosch algorithm~\cite{bk1973}.
The algorithm involves traversing a search tree that branches from parent nodes representing smaller cliques to child nodes representing larger cliques that contain them until maximal cliques are found.
Prior works on parallelizing MCE on GPUs perform a breadth-first traversal of the search tree~\cite{mcg3, mcg1, mcg2, mcg6_2021}.
However, this approach does not scale well for large graphs because of the explosion in the number of search tree nodes that need to be tracked at deep levels of the tree.
To overcome this limitation, we propose to parallelize MCE on GPUs by assigning independent subtrees to different thread blocks and having the threads within each block collaboratively perform a depth-first traversal of the block's subtree.

The approach of performing per-block depth-first traversal of independent subtrees has been applied in our prior work on $k$-clique counting~\cite{ourkclique}.
However, MCE presents two key scalability challenges that are less of a concern in $k$-clique counting.
The first challenge is that the MCE search tree is substantially more imbalanced, which means that assigning independent subtrees to different thread blocks suffers from high load imbalance.
The second challenge is that MCE requires substantially more memory capacity to track the vertices excluded at each level of the traversal to test for maximality of a clique.
These two challenges are particularly critical on GPUs in contrast with CPUs.
GPUs are more sensitive to load imbalance than CPUs due to their massively parallel nature~\cite{wenmeibook}.
Moreover, GPUs typically have a smaller memory capacity than CPUs while putting more pressure on the memory capacity by traversing a larger number of subtrees in parallel.

In this paper, we propose a novel solution for accelerating MCE on GPUs that employs various techniques to address the load imbalance and memory capacity challenges that MCE imposes.
We propose a worker list to enable thread blocks with large subtrees to offload branches of their subtrees to other thread blocks with low overhead.
We propose using partial induced subgraphs to avoid the latency and memory capacity overhead of constructing full induced subgraphs.
We propose a compact representation of the sets of excluded vertices that distinguishes between the part of each set that needs to be stored separately for each level, and the part that monotonically shrinks and can be reused across levels.
We also retain several optimizations used in our prior work on $k$-clique counting~\cite{ourkclique}, such as binary encoding of the induced subgraph and partitioning work at subwarp granularity.

Our evaluation shows that our parallel GPU implementation executing on a single server-grade GPU outperforms the state-of-the-art parallel CPU implementation~\cite{mcc6_baseline} executing on a server-grade CPU by a geometric mean of 4.9$\times$ (up to 16.7$\times$).
We also show that our worker list approach is effective at achieving load balance with low overhead, and enables efficient scaling to multiple GPUs.
Our code has been open-sourced for reproducibility and to enable further research on accelerating MCE.

%% file: sec/2-background.tex
\section{Background}

\subsection{Maximal Clique Enumeration}

\begin{figure*}
    \centering
    \includegraphics[width=\textwidth]{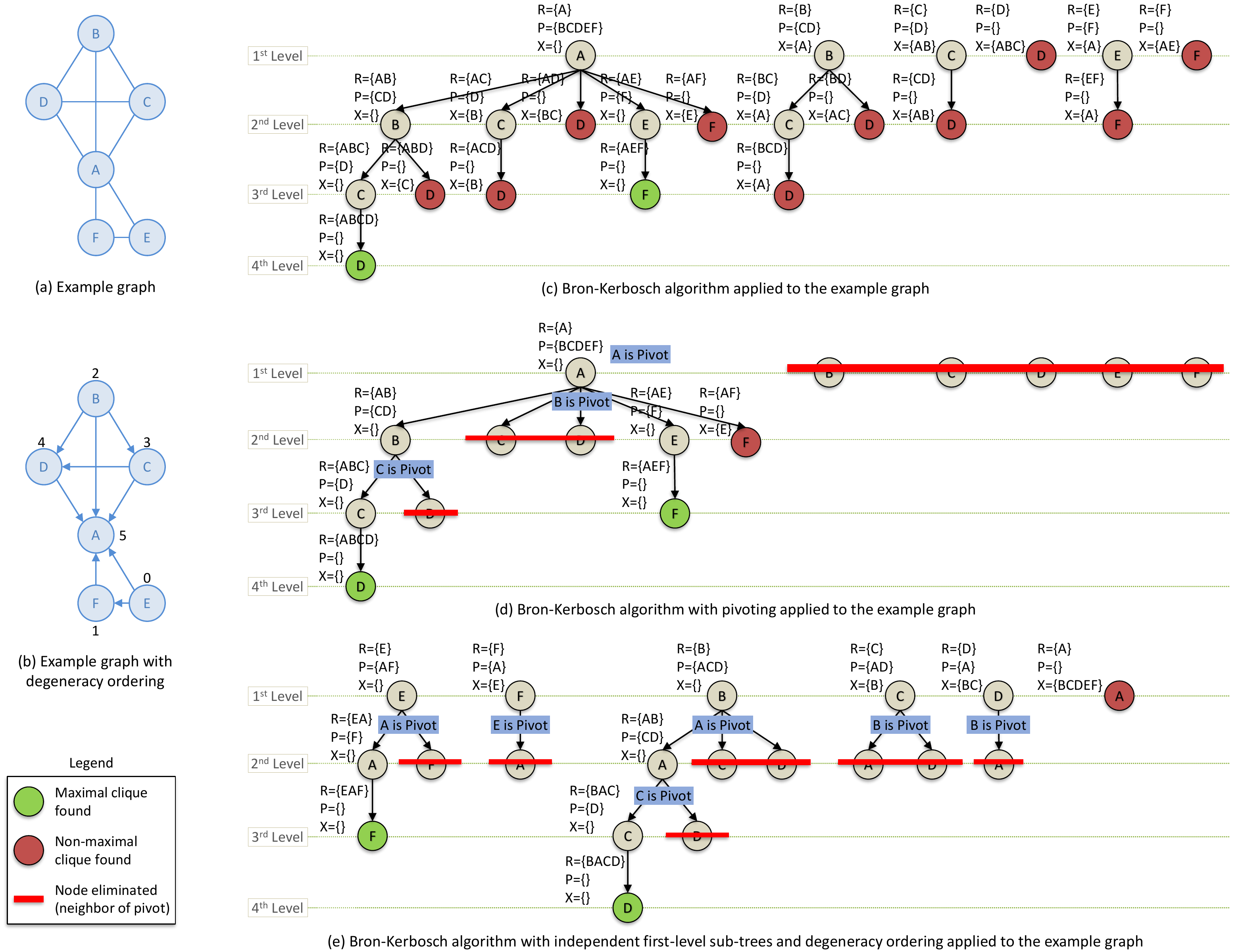}
    \caption{Bron-Kerbosch algorithm variants applied to the example graph}\label{fig:bk-example}
\end{figure*}

Let $G = (V, E)$ be a simple undirected graph where $V$ is the set of vertices in $G$ and $E$ is the set of edges in $G$.
The \textit{neighborhood} of a vertex $v \in V$ is the set of vertices adjacent to $v$, denoted by $N(v)$.
A \textit{clique} in $G$ is a complete subgraph of $G$ where every vertex in the subgraph is adjacent to every other vertex in the subgraph.
A \textit{maximal clique} in $G$ is a clique that cannot be further expanded by including one more vertex.
In other words, a maximal clique is a clique that is not contained in a larger clique.
For example, the graph in Fig.~\ref{fig:bk-example}(a) has two maximal cliques: ABCD and AEF.
On the other hand, ABC, ABD, ACD, and BCD are not maximal cliques because they are all contained in ABCD.

MCE aims to find all the maximal cliques in a graph.
We tackle MCE as an exact problem, which means that we enumerate all maximal cliques in the graph and do not apply any approximations or graph sampling.
While approximate maximal cliques may be sufficient for some applications, other applications require exact maximal cliques. For example, in protein-protein interaction networks, a protein complex is necessarily a clique~\cite{gagneur2004modular}.
In addition, even for applications where approximate solutions can be used, there is added value in using an exact solution if finding it can be made sufficiently efficient.

One of the most widely used algorithms for exact MCE is the Bron-Kerbosch algorithm~\cite{bk1973}.
We describe different variants and optimizations of this algorithm in the rest of this section.

\subsection{Bron-Kerbosch}\label{sec:bk}

The Bron-Kerbosch algorithm~\cite{bk1973} is a backtracking algorithm that traverses a search tree to find maximal cliques.
The search tree branches from parent nodes representing smaller cliques to child nodes representing larger cliques that contain them until maximal cliques are found.
While searching, the algorithm maintains three disjoint sets for each tree node: result ($R$), possible ($P$), and exclude ($X$).
$R$ is the set of vertices in the clique currently being explored.
$P$ and $X$ together contain the common neighbors of the vertices in $R$.
$P$ is the set of common neighbors that can still be added to the clique in $R$ in the current branch of the tree.
$X$ is the set of common neighbors that have already been considered in another branch of the tree, so they are excluded from the maximal clique being searched for in the current branch.

Algorithm~\ref{alg:bk1} shows the pseudocode for the Bron-Kerbosch algorithm, and Fig.~\ref{fig:bk-example}(c) shows how this algorithm is applied to the example graph in Fig.~\ref{fig:bk-example}(a).
In the initial call to \textsc{BronKerbosch}, $R$ is empty, $P$ contains all the vertices in the graph, and $X$ is empty.
The recursive step (lines 5-8) iterates over all the vertices $v$ in the set $P$.
At the recursive call (line 6), $v$ is added to the solution $R$, and $P$ and $X$ are intersected with $N(v)$ to remove non-neighbors of $v$.
After returning from the call, all maximal cliques containing the vertices in $R \cup \{v\}$ have been found.
To avoid finding the same cliques again, $v$ is excluded from the search in later subtrees at the same level by removing $v$ from $P$ (line 7) and adding to $X$ (line 8) before proceeding to the next loop iteration.

\begin{algorithm}[ht]
    \small
    \caption{Bron-Kerbosch algorithm}\label{alg:bk1}
    \begin{algorithmic}[1]
        \input{fig/02-background/bk1}
    \end{algorithmic}
\end{algorithm}

The recursion stops when $P$ is empty, which means that there are no more vertices that can be added to the clique in $R$.
If $P$ and $X$ are both empty (line 2), then the vertices in $R$ have no common neighbors, which means that $R$ represents a maximal clique (line 3).
If $P$ is empty but $X$ is not empty, then the vertices in $R$ do have common neighbors (those in $X$) and $R$ is not a maximal clique.
However, the search stops because the common neighbors in $X$ are excluded from the search on this tree branch, which means that any maximal clique containing $R$ has already been found in other branches.

\subsection{Bron-Kerbosch with Pivoting}\label{sec:bk-pivot}

It is clear from Fig.~\ref{fig:bk-example}(c) that there can be many branches in the search tree that are not successful at finding a maximal clique because the clique is found by other branches.
To avoid some of the unsuccessful branches, Bron and Kerbosch introduce \textit{pivoting}~\cite{bk1973}.
Algorithm~\ref{alg:bk2} shows the pseudocode for the Bron-Kerbosch algorithm with pivoting, and Fig.~\ref{fig:bk-example}(d) shows how this algorithm is applied to the example graph in Fig.~\ref{fig:bk-example}(a).
The difference from Algorithm~\ref{alg:bk1} is that in Algorithm~\ref{alg:bk2}, a \textit{pivot vertex} is selected prior to branching (line 5) and the neighbors of the pivot vertex are excluded from the branching (line 6).

\begin{algorithm}[ht]
    \small
    \caption{Bron-Kerbosch algorithm with pivoting}\label{alg:bk2}
    \begin{algorithmic}[1]
        \input{fig/02-background/bk2}
    \end{algorithmic}
\end{algorithm}

The intuition behind pivoting is that any maximal clique that includes the pivot vertex and its neighbor will be found by the branch that adds the pivot vertex to $R$.
On the other hand, any maximal clique that does not include the pivot vertex but includes its neighbor must include a non-neighbor of the pivot vertex, and will be found by the branches that add non-neighbors of the pivot vertex to $R$.
Therefore, there is no need to explore the pivot vertex's neighbors on line 6.

The pivot vertex can be any vertex in $P \cup X$, but is typically selected to have the largest number of neighbors that are also in $P$ in order to maximize the number of branches that are excluded from the search.
The original Bron-Kerbosch algorithm with pivoting selects the pivot vertex from $P$, but Tomita et al.~\cite{bk_tomita} improve the pivot selection by considering all vertices in $P \cup X$.

\subsection{Bron-Kerbosch with Other Optimizations}\label{sec:bk-pivot-order}

Eppstien et al.~\cite{bk_eppstein, arbolow1} further improve the Bron-Kerbosch algorithm with pivoting by introducing three key optimizations: independent first-level subtrees, degeneracy ordering, and induced subgraphs.

\textbf{Independent First-level Subtrees.}
In Algorithms~\ref{alg:bk1} and~\ref{alg:bk2}, the loop that iterates over the vertices in $P$ has a loop-carried dependence for removing previously visited vertices from $P$ and adding them to $X$.
Eppstien et al.~\cite{bk_eppstein, arbolow1} break this loop-carried dependence at the first level by having each iteration independently remove all previous vertices from $P$ and add them to $X$.
The pseudocode for doing so is shown in Algorithm~\ref{alg:bk3}.
In each iteration, $P$ for vertex $v_i$ is initialized by intersecting $N(v_i)$ with the set of vertices that come after $v_i$ (line 3), which removes the neighbors of $v_i$ visited on prior iterations.
On the other hand, $X$ for vertex $v_i$ is initialized by intersecting $N(v_i)$ with the set of vertices that come before $v_i$ (line 4), which \textit{keeps} the neighbors of $v_i$ visited on prior iterations.
The advantage of breaking this loop-carried dependence is that the loop iterations, which represent first-level subtrees, can be executed in parallel.
For the second level onward, the algorithm simply calls the sequential \textsc{BronKerboschPivot} function (line 5).

\begin{algorithm}[ht]
    \small
    \caption{Bron-Kerbosch algorithm with independent first-level subtrees}\label{alg:bk3}
    \begin{algorithmic}[1]
        \input{fig/02-background/bk3}
    \end{algorithmic}
\end{algorithm}

\textbf{Degeneracy Ordering.}
In Algorithm~\ref{alg:bk3}, the subtree for each vertex $v_i$ only considers the vertices in $P$.
Moreover, in Algorithm~\ref{alg:bk2} (which is called by Algorithm~\ref{alg:bk3} on line 5), several expensive set operations are performed with $P$ such as $P-N(v_{pivot})$ (line 6) and $P \cap N(v)$ (line 7).
Hence, the size of $P$ directly impacts the size of the subtree traversed and the cost of the set operations performed by the traversal.
Recall that $P$ represents the neighbors of $v_i$ that are ordered after $v_i$, and $X$ represents the neighbors of $v_i$ that are ordered before $v_i$.
For an arbitrary graph, the sizes of $P$ and $X$ are $O(\Delta)$ where $\Delta$ denotes the maximum degree of the graph and can be quite large for real graphs.
To place a tighter bound on the size of $P$, Eppstien et al.~\cite{bk_eppstein} propose to reorder vertices based on \textit{degeneracy ordering} which minimizes the maximum number of neighbors of any vertex that are ordered after that vertex.
After degeneracy ordering, the maximum number of neighbors of any vertex that are ordered after that vertex is known as the degeneracy of the graph, and is denoted by $d$.
The size of $P$ thus becomes $O(d)$.
For real graphs, $d$ is typically much smaller than $\Delta$ (see Table~\ref{tab:dataset}).

The advantage of degeneracy ordering is that by placing a smaller bound on the sizes of the $P$ sets, it places a smaller bound on the sizes of the subtrees traversed and the cost of the set operations performed with $P$.
However, the size of the $X$ sets remains $O(\Delta)$, and the practical size of the maximum $X$ set increases due to degeneracy ordering.
Hence, the trade-off of degeneracy ordering is that it makes the operations on the $X$ sets, such as $X \cap N(v)$ (line 7 in Algorithm~\ref{alg:bk2}), more expensive.

In Fig.~\ref{fig:bk-example}(a), the first vertex $A$ was also the vertex with the highest degree, which resulted in large subtrees being visited for vertex $A$ in Fig.~\ref{fig:bk-example}(c) and Fig.~\ref{fig:bk-example}(d).
The size of the $P$ set for $A$ was five which is the maximum degree of the graph.
Fig.~\ref{fig:bk-example}(b) shows how the graph in Fig.~\ref{fig:bk-example}(a) can be reordered based on degeneracy ordering.
In this figure, the graph is still intended to be undirected, but the edges are drawn with arrows from vertices earlier in the order to vertices later in the order.
As shown in Fig.~\ref{fig:bk-example}(b), $A$ is now the last vertex in the order and has no vertices ordered after it.
Fig.~\ref{fig:bk-example}(e) shows how the example graph in Fig.~\ref{fig:bk-example}(a) can be processed using independent first-level subtrees and degeneracy ordering.
It is clear that compared to Fig.~\ref{fig:bk-example}(c) and Fig.~\ref{fig:bk-example}(d), Fig.~\ref{fig:bk-example}(e) has more independent subtrees that are each smaller in size, its largest $P$ set is smaller, and its largest $X$ set is larger.

\textbf{Induced Subgraphs.}
In Algorithm~\ref{alg:bk3}, the subtree for each vertex $v_i$ only needs to access the neighbors of $v_i$ and their edges.
It does not need to access the entire graph.
Based on this observation, Eppstien et al.~\cite{bk_eppstein} propose to construct an induced subgraph for each subtree that only includes the information needed by that subtree.
In particular, we observe that Algorithm~\ref{alg:bk2} performs three key operations that access the graph.
The first operation is $P-N(v_{pivot})$ (line 6).
Since $v_{pivot} \in P \cup X$, this operation needs to know the neighbors of any vertex in $P \cup X$ that are in $P$.
The second operation is $P \cap N(v)$ (line 7).
Since $v \in P$, this operation needs to know the neighbors of any vertex in $P$ that are also in $P$.
The third operation is $X \cap N(v)$ (line 7).
Since $v \in P$, this operation needs to know the neighbors of any vertex in $P$ that are in $X$.
Overall, the algorithm needs the edges connecting any vertex in $P \cup X$ with any vertex in $P$.
Eppstien et al.~\cite{bk_eppstein} induce a subgraph that contains only this information, denoted by $H_{P,X}$.
The key advantage of using an induced subgraph is that it removes irrelevant edges from the adjacency lists, making set operations on the adjacency lists smaller.

Without degeneracy ordering, the size of $P \cup X$ is $O(\Delta)$ and the size of $P$ is $O(\Delta)$.
Hence, the size of $H_{P,X}$ is $O(\Delta^2)$ which is prohibitively expensive to store for large graphs.
However, after degeneracy ordering, the size of $P$ is reduced to $O(d)$, which reduces the size of $H_{P,X}$ to $O(\Delta \cdot d)$.
Since typically $d \ll \Delta$, degeneracy ordering makes it more feasible to construct and store an induced subgraph.

%% file: fig/02-background/bk1.tex
\Procedure{BronKerbosch}{$G$, $R$, $P$, $X$}

    \If{$P$ and $X$ are both empty} 
        \State $R$ is a maximal clique
        \State \Return
    \EndIf

    \For{$v \in P$}
        \State \textsc{BronKerbosch}($G$, $R \cup \{v\}$, $P \cap N(v)$, $X \cap N(v)$)
        \State $P = P - \{v\}$
        \State $X = X \cup \{v\}$
    \EndFor
\EndProcedure

%% file: fig/02-background/bk2.tex
\Procedure{BronKerboschPivot}{$G$, $R$, $P$, $X$}
    \If{$P$ and $X$ are both empty}
        \State $R$ is a maximal clique
        \State \Return
    \EndIf
    \State $v_{pivot}$ = choose a vertex from $P \cup X$
    \For{$v \in (P - N(v_{pivot}))$}
        \State \textsc{BronKerboschPivot}($G$, $R \cup \{v\}$, $P \cap N(v)$, $X \cap N(v)$)
        \State $P = P - \{v\}$
        \State $X = X \cup \{v\}$
    \EndFor
\EndProcedure

%% file: fig/02-background/bk3.tex
\Procedure{BronKerboschIndependentFirstLevel}{$G$}
     \For{$v_i \in V$}
         \State $P = N(v_i) \cap \{v_{i+1}, v_{i+2}, ..., v_{|V|-1}\}$
         \State $X = N(v_i) \cap \{v_{0}, v_{1}, ..., v_{i-1}\}$
         \State \textsc{BronKerboschPivot}($G$, $\{v_i\}$, $P$, $X$)
     \EndFor
\EndProcedure

%% file: sec/3-approach.tex
\section{Parallelizing MCE on GPUs}

\subsection{Challenges and Implementation Overview}\label{sec:challenges}

We propose a parallel implementation of MCE on GPUs based on the Bron-Kerbosch algorithm with pivoting, independent first-level subtrees, degeneracy ordering, and induced subgraphs.
One of the main challenges for parallelizing the Bron-Kerbosch algorithm on GPUs is extracting a sufficient amount of parallelism to fully-utilize the hardware resources.
The majority of prior works~\cite{mcg3, mcg1, mcg2, mcg6_2021} do so by performing a breadth-first traversal of the search tree.
Breadth-first search is highly amenable to parallelization because tree nodes at each level of the search tree can be processed in parallel.
However, it does not scale well for large graphs because of the explosion in the number of search tree nodes that need to be tracked as the level gets deeper.
To avoid this explosion, one work~\cite{lessons} performs depth-first search on CPU while offloading primitive operations to GPU.
However, this approach results in high communication overhead between CPU and GPU due to frequent kernel calls and data transfer operations.

To overcome these limitations, we propose to parallelize MCE on GPUs by assigning independent subtrees to different thread blocks and having each thread block perform a depth-first traversal of its subtree.
Threads within the block collaborate to perform primitive operations such as set operations and finding pivots.
This approach prevents the explosion in the number of search tree nodes that need to be tracked, and performs the entire traversal in a single kernel which eliminates CPU-GPU communication.
There is no communication between CPU and GPU throughout the execution, except copying the graph to the GPU at the beginning and copying the result back at the end.

The parallelization approach of performing per-block depth-first traversals of independent subtrees has been applied in our prior work on $k$-clique counting~\cite{ourkclique}.
That work also applies other optimizations such as binary encoding of the induced subgraph and partitioning work within a block at subwarp granularity.
In this work, we retain all these optimizations.
To the best of our knowledge, this work is the first to use induced subgraphs, binary encoding, and subwarp partitioning for parallelizing MCE on GPUs.

Aside from applying these techniques to MCE, our main contribution in this work is addressing two key scalability challenges present in MCE that are less of a concern in $k$-clique counting.
The first challenge is that MCE has substantially higher load imbalance.
In $k$-clique counting, search trees have bounded depth (i.e., $k$).
Hence, the sizes of subtrees that are assigned to different thread blocks are reasonably balanced.
Moreover, our prior work on $k$-clique counting~\cite{ourkclique} shows that extracting independent subtrees at the second level instead of the first level is sufficient to balance the load completely.
In contrast, in MCE, the subtrees may be arbitrarily deep depending on the size of the maximal clique they are exploring.
Hence, MCE suffers from substantially higher load imbalance than $k$-clique counting and requires more sophisticated load balancing techniques.

The second challenge is that MCE has a substantially higher memory footprint than $k$-clique counting.
Since MCE has potentially deeper subtrees, it needs to pre-allocate more stack space per thread block to support the depth-first traversal of these subtrees.
Moreover, in $k$-clique counting, the traversal only needs to track the equivalent of the $R$ and $P$ sets at each level of the tree, which are $O(d)$ in size, and the induced subgraphs only need to store the edges between vertices in $P$ and other vertices also in $P$, which requires $O(d^2)$ space.
In contrast, in MCE, to test for maximality, the traversal also needs to track the $X$ set for each level of the tree, which is $O(\Delta)$ in size, and the induced subgraphs also need to store the edges between vertices in $X$ and vertices in $P$, which requires $O(\Delta \cdot d)$ space.
Since $\Delta$ is much larger than $d$, MCE has a substantially higher memory footprint than $k$-clique counting and requires more sophisticated techniques for representing induced subgraphs and the $X$ sets.

In the rest of this section, we describe our proposed approach for parallelizing MCE on GPUs, with a particular focus on unique aspects of our work, namely, how to mitigate load imbalance and how to efficiently represent induced subgraphs and the $X$ sets at each level of the search tree.

\subsection{Independent Second-level Subtrees}\label{sec:l2-trees}

One common approach to improving load balance on GPUs is to extract many more parallel tasks than the number of tasks that can be executed simultaneously by the hardware.
Our prior work on $k$-clique counting~\cite{ourkclique} advocates for extracting independent subtrees at the second level instead of the first, and shows that it is sufficient to balance load completely for that problem.
We investigate the same technique in MCE.

Algorithm~\ref{alg:bk4} shows the pseudocode for extracting independent second-level subtrees.
Instead of iterating over vertices in $V$, we iterate over edges $\{v_i, v_j\}$ in $E$ (line 2).
For each edge, $P$ is initialized to the common neighbors of $v_i$ and $v_j$ that are ordered after both vertices (line 3).
On the other hand, $X$ is initialized to the common neighbors of $v_i$ and $v_j$ that are ordered before the later of the two vertices (line 4).

\begin{algorithm}[ht]
    \small
    \caption{Bron-Kerbosch algorithm with independent second-level subtrees}\label{alg:bk4}
    \begin{algorithmic}[1]
        \input{fig/02-background/bk4}
    \end{algorithmic}
\end{algorithm}

The advantage of extracting subtrees at the second level instead of the first level is that it provides more parallel tasks to assist with load balancing.
It also results in smaller induced subgraphs since the $P$ sets at the second level are smaller than those at the first level.
The disadvantage is that more induced subgraphs need to be constructed overall, and their construction cost is amortized across smaller subtree traversals.

We evaluate the trade-off between extracting subtrees at the first or second level throughout Section~\ref{sec:eval}.
We observe that although extracting second-level subtrees partially reduces load imbalance, the imbalance remains high for many graphs unlike in $k$-clique counting.
This observation motivates us to propose another optimization for mitigating load imbalance in MCE, which is more effective and ultimately obviates the need for extracting second-level subtrees.

\subsection{Dynamic Load Balancing with a Worker List}\label{sec:wl}

One approach to alleviate load imbalance on GPUs is using a \textit{worklist}.
Thread blocks with large tasks can add subtasks to the worklist, and thread blocks that complete their tasks can remove subtasks from the worklist.
For example, Yamout et al.~\cite{yamout2022parallel} use such an approach to achieve load balance while traversing the vertex cover search tree, leveraging the broker worker distributor~\cite{kerbl2018broker} as their worklist data structure.
However, in MCE, the data needed to represent a subtask is large, consisting of $R$, $P$, $X$, the current level, and a reference to the induced subgraph.
The large size of the subtask data makes using a worklist inefficient for MCE for two reasons.
The first reason is that a large amount of memory would be needed to store the worklist entries, which would place high pressure on the already constrained memory capacity.
The second reason is that adding and removing subtasks from the worklist would incur high overhead, so there would be a high penalty when a block adds work to a worklist and there are no idle blocks actually needing any work.

To avoid the limitations of using a worklist, we instead propose to use a \textit{worker list} for dynamic load balancing.
The worker list holds IDs of thread blocks that are idle because they have completed their previous tasks.
A thread block that completes its task adds its ID to the worker list to indicate that it can receive subtasks from other blocks.
We call this block a \textit{receiver} block.
On the other hand, a thread block working on a large task periodically checks the worker list to see if there are any receiver blocks waiting.
We call this block a \textit{donor} block.
If a donor block finds a receiver block in the worker list, the donor block removes the receiver block and gives it a subtask to work on.
The computation terminates when all blocks have added themselves to the worker list and there are no more executing donor blocks.

We incorporate our proposed worker list technique into our parallel MCE implementation as follows.
We start by launching as many thread blocks as the maximum number that can run on the GPU simultaneously.
These blocks execute in two phases.
In the first phase, each block atomically increments a shared counter to reserve an independent first- or second-level subtree, and traverses that subtree.
If the block completes the subtree, it atomically increments the counter again to obtain another subtree.
This process continues until all the independent subtrees have been depleted, after which the second phase begins.
Note that there is no global synchronization needed between the two phases.
Donor blocks that are still executing their subtrees from the first phase know when the second phase has been reached by checking the shared counter every time they branch.

In the second phase, blocks that finish traversing their subtrees add themselves to the worker list and sleep by spinning on a flag with exponential back-off.
The worker list is implemented as a multi-producer multi-consumer queue using a circular buffer.
The buffer cannot overflow because the number of thread blocks is fixed.
Donor blocks that have not finished traversing their subtrees check the worker list upon visiting a new branch.
If a donor finds a receiver in the worker list, the donor atomically removes the receiver's ID from the worker list, offloads the new branch to the receiver by initializing the receiver's data structures, and wakes the receiver up by setting its flag.
We use CUDA atomic objects from libcu++ and the release-acquire model to guarantee that data written by the donor is visible to the receiver.

In some cases, the benefit of a donor block offloading a branch to a receiver block is not worth the overhead.
To avoid these unprofitable cases, we only have a donor block check the worker list and offload a branch if two conditions hold.
The first condition is that the branch to be offloaded should not be small, otherwise the overhead of offloading the branch to the receiver may be higher than the cost of visiting the branch.
To ensure that the branch is not small, we require that $|P| \geq 10$ for the root node of the branch, however we note that performance is not very sensitive to the choice of this threshold.
The second condition is that the donor block should have a substantial amount of other work to do after offloading the branch, because it does not make sense for the donor to offload a branch, then finish traversing its subtree shortly after and start seeking work from other donors.
To ensure that the donor has a substantial amount of other work, we only offload a branch if there are other branches at the same level \textit{and} other branches in previous levels that have not yet been explored.

We evaluate the advantage of using a worker list in Sections~\ref{sec:eval-load-balance} and~\ref{sec:eval-scalability}, including its importance when scaling to multiple GPUs.

\subsection{Partial Induced Subgraphs}\label{sec:induced}

Recall from Section~\ref{sec:bk-pivot-order} that one common optimization to reduce the size of adjacency lists and intersection operations is to construct, for each independent subtree, an induced subgraph with vertices and edges relevant to that subtree.
Prior works that implement MCE on GPUs~\cite{mcg3, mcg1, mcg2, mcg6_2021,lessons} do not apply this optimization because they do not perform depth-first traversal of independent subtrees entirely on the GPU.
To the best of our knowledge, our work is the first to use induced subgraphs for MCE on GPUs.

As mentioned in Section~\ref{sec:challenges}, the induced subgraphs in MCE contain the edges between the vertices in $P$ and the vertices in $P \cup X$, which makes the size of the induced subgraph $O(\Delta \cdot d)$.
Since $\Delta$ can be large, these induced subgraphs are expensive to construct and store.
To address this challenge, we propose to represent the induced subgraph using two alternatives: full or partial.
For full induced subgraphs, we construct binary-encoded induced subgraphs containing all the edges between $P$ and $P \cup X$.
For partial induced subgraphs, we construct binary-encoded induced subgraphs with only the edges between vertices $P$ and other vertices also in $P$, and use the original graph to look up edges between vertices in $P$ and vertices in $X$.
The original graph is stored using the Compressed Sparse Row (CSR) format~\cite{wenmeibook} when first-level subtrees are used, and both the CSR and the Coordinate format (COO)~\cite{wenmeibook} when second-level subtrees are used.

The advantage of using full induced subgraphs is that it makes set operations on $X$ faster by using bitwise operations.
The advantage of using partial induced subgraphs is that it avoids the high latency of constructing large induced subgraphs and the high memory capacity required for storing them.
We evaluate these trade-offs and propose a heuristic for selecting the most suitable alternative in Section~\ref{sec:eval-breakdown}.

\subsection{Compact Representation of the \texorpdfstring{$X$}{X} Sets}

Recall from Section~\ref{sec:challenges} that MCE puts higher pressure on the memory capacity than other related problems because of the need to represent $X$ at each level of a subtree to test for maximality.
A subtree can have up to $d$ levels, and the size of $X$ is $O(\Delta)$.
Therefore, a naive representation of the $X$ sets would require $O(\Delta \cdot d)$ memory per subtree.
Hence, the memory needed to represent the $X$ sets can easily limit the number of subtrees that can be traversed in parallel.

To design an efficient representation of $X$, we first make the following observations.
In Algorithm~\ref{alg:bk2}, the two operations that modify $X$ as the tree is traversed are $X~\cap~N(v)$ (line~7) and $X~\cup~\{v\}$ (line 9), where $v \in P$.
The first operation, $X \cap N(v)$, can only remove vertices from $X$.
The second operation, $X~\cup~\{v\}$, adds vertices to $X$ but these vertices can only come from $P$.
Based on this observation, we divide the representation of $X$ into two parts: $X_P$ and $X_X$.

$X_P$ represents the vertices in $X$ that are part of the original $P$ set at the root node of the subtree.
These vertices may be added by the $X~\cup~\{v\}$ operation or removed by the $X \cap N(v)$ operation.
Hence, $X_P$ may grow or shrink as we descend to deeper levels of the subtree.
However, $X_P$ may not exceed the size of $P$ which is $O(d)$.
For this reason, $X_P$ is binary encoded for fast set operations on it, and a different copy of $X_P$ is stored for each level of the tree.

On the other hand, $X_X$ represents the vertices in $X$ that were part of the original $X$ set at the root node of the subtree.
$X_X$ may contain any vertex in the original $X$ which makes its size $O(\Delta)$.
However, since the vertices in the original $X$ cannot be part of any $P$ set in the subtree, the vertices in $X_X$ may only be removed by the $X \cap N(v)$ operation as we descend to deeper levels of the subtree.
Since $X_X$ only shrinks as we descend down the subtree, we do not need to store a separate copy of $X_X$ for each level.
Instead, we store a single copy of $X_X$ for all levels and an index for each level that points to where the $X_X$ vertices end for that level.

\begin{figure}[t]
    \centering
    \includegraphics[width=0.8\columnwidth]{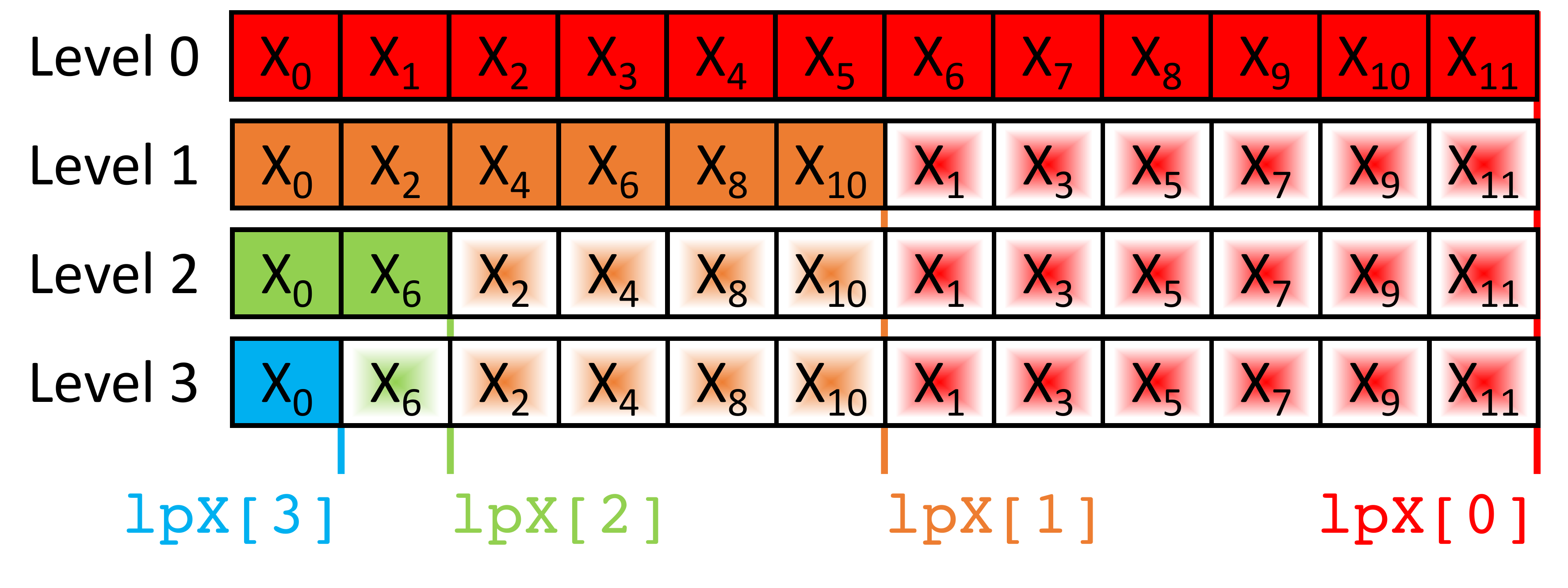}
    \caption{Using a single array to represent $X_X$ across levels}
    \label{fig:mce_shrinking_x}
\end{figure}

Fig.~\ref{fig:mce_shrinking_x} shows an example of how $X_X$ is represented and updated as we descend down the tree.
In this example, as we descend from Level 0 to Level 1, the vertices that remain in $X$ in Level 1 are moved to the front of the array and the vertices that are removed are moved to the end of the array.
To move the vertices, we implement an out-of-place partition operation where each thread moves one value after atomically incrementing a bin counter.
We also tried the in-place partition operation in CUB~\cite{cub} but it did not yield better performance.
After moving the vertices, a level pointer array \texttt{lpX} is updated such that \texttt{lpX[1]} points to where the vertices in Level 1 end.
The same process is repeated on the shrunk array as we descend to deeper levels.
To go back to a previous level, nothing needs to be done since all the vertices for the previous level have stayed before the previous level's \texttt{lpX} pointer and only the order of vertices has changed.

By storing a single copy of $X_X$ for all levels and different copies of $X_P$ for each level, the memory needed for representing $X$ for all levels becomes $O(\Delta + d^2)$ which is much smaller than $O(\Delta \cdot d)$.
This compact representation is crucial for scalable acceleration of MCE on GPUs (and any other memory constrained system) and is used in all our implementations.

Finally, we note that if a partial induced subgraph is used instead of a full induced subgraph (see Section~\ref{sec:induced}), then the pivot is only selected from $X_P \cup P$.
The reason is that finding a pivot vertex from $X_X$ is expensive if the adjacency lists of the vertices in $X_X$ are not binary encoded.

%% file: fig/02-background/bk4.tex
\Procedure{BronKerboschIndependentSecondLevel}{$G$}
     \For{$\{v_i, v_j\} \in E$}
         \State $P = N(v_i) \cap N(v_j) \cap \{v_{max(i,j)+1}, ..., v_{|V|-1}\}$
         \State $X = N(v_i) \cap N(v_j) \cap \{v_{0}, ..., v_{max(i,j)-1}\}$
         \State \textsc{BronKerboschPivot}($G$, $\{v_i, v_j\}$, $P$, $X$)
     \EndFor
\EndProcedure

%% file: sec/4-evaluation.tex
\section{Evaluation}\label{sec:eval}

\begin{table*}[t]
    \centering
    \caption{Graphs used for evaluation and comparison of execution time with the state-of-the-art parallel CPU implementation}\label{tab:dataset}
    \resizebox{\textwidth}{!}{
        \input{fig/04-evaluation/tab-dataset}

    }
\end{table*}

\subsection{Methodology}\label{sec:methodology}

\textbf{Evaluation Platforms.}
We evaluate our GPU implementations on two platforms.
The first platform has four 32GB NVIDIA V100 GPUs attached to an Intel Xeon Gold 6230 CPU and is used for both single- and multi-GPU evaluation.
On this platform, we compile our code with NVCC (CUDA 10.2) and GCC 8.3.1 with the -O3 flag.
The CUDA driver version is 470.74.
The second platform has a 40GB NVIDIA A100 GPU attached to an AMD EPYC 7702 CPU and is used for the single-GPU evaluation only.
On this platform, we compile our code with NVCC (CUDA 11.4) and GCC 9.4.0 with the -O3 flag.
The CUDA driver version is 470.103.
We use 128 threads per block, which results in 1,280 blocks per GPU for V100 and 1,728 blocks per GPU for A100.
We use CUB 1.8.0~\cite{cub} for the filter and exclusive scan operations during pre-processing.
In the multi-GPU implementations, we use OpenMP 4.5 to create one CPU thread for each GPU.

\textbf{CPU Baseline.}
To the best of our knowledge, the work of Blanuša et al.~\cite{mcc6_baseline} is the state-of-the-art parallel CPU implementation of MCE, and also outperforms all prior GPU implementations.
We compare the performance of our GPU implementation with the best execution times reported by Blanuša et al. which are obtained using a dual-socket Intel Xeon Skylake platform with 48 cores (96 threads) and 360 GB of main memory.
For completeness, we also execute their publicly available code on our dual-socket Intel Xeon Gold 6230 Cascade Lake CPU with 40 cores (80 threads) and 512GB of main memory and report those results as well.

\textbf{GPU Baseline.}
Prior GPU implementations do not have publicly available code.
For this reason, we compare the performance of our implementation to the execution times reported in the most recent GPU work by Wei et al.~\cite{mcg6_2021}.
However, this comparison is not fair because Wei et al. use an NVIDIA Titan X GPU which is weaker than the GPUs we use.
We comment on this issue in Section~\ref{sec:eval-performance}.

\textbf{Datasets}
We evaluate using the same graph datasets used by Blanuša et al.~\cite{mcc6_baseline} which are shown in Table~\ref{tab:dataset}.

\textbf{Reporting of Measurements.}
The execution times reported by Blanuša et al.~\cite{mcc6_baseline} include the time spent on counting maximal cliques and exclude the time spent on reading the graph from disk.
For fair comparison, we follow the same strategy.
We also include the time spent on pre-processing the graph to apply degeneracy ordering.
Unless otherwise specified, we report the time achieved with the worker list enabled, and with the best combination of using independent first- or second-level subtrees and using partial or full induced subgraphs.

\subsection{Performance}\label{sec:eval-performance}

\textbf{Performance comparison with prior CPU implementation.}
Table~\ref{tab:dataset} compares the execution time of our single-GPU implementation with the state-of-the-art parallel CPU implementation~\cite{mcc6_baseline}.
We observe that our GPU implementation consistently and significantly outperforms the parallel CPU implementation for all graphs.
The geometric mean speedup of our GPU implementation over the parallel CPU implementation is 4.1$\times$ (up to 10.2$\times$) for the V100 GPU and 4.9$\times$ (up to 16.7$\times$) for the A100 GPU.
These results show the effectiveness of GPUs at accelerating MCE, despite the challenges of GPUs being more sensitive to load imbalance and having more constrained memory capacity.
Note that while some of the optimizations introduced in this paper may be applied to CPU implementations, we do not expect them to be as effective because CPU implementations do not suffer as much from load imbalance and memory capacity constraint.

\textbf{Performance comparison with prior GPU implementation.}
Table~\ref{tab:gpu-baseline} compares the execution time of our single-GPU implementation with the most recent GPU implementation~\cite{mcg6_2021} for the common graphs reported in that implementation.
The geometric mean speedup of our GPU implementation over the prior GPU implementation is 35.65$\times$ (up to 50.46$\times$).
As mentioned in Section~\ref{sec:methodology}, this comparison is not fair because Wei et al. use a Titan X GPU which is weaker than our V100 GPU.
However, the V100 GPU has only 1.43x more cores and 1.88x higher memory bandwidth than the Titan X GPU so the achieved speedup cannot be attributed to the hardware difference alone.
Unfortunately, we are unable to make a direct comparison on the same system because we do not have access to an NVIDIA Titan X GPU to evaluate our implementation on, nor do we have access to Wei et al.'s code to evaluate it on our system.

\textbf{Performance relative to hardware peak capability.}
The efficiency of our implementation relative to the hardware peak capability depends on the graph being solved.
For many graphs, the computation is compute-bound (computing set intersections) when the induced subgraph fits in the L1 cache.
It shifts towards memory-bandwidth-boundedness when the induced subgraph is large and global memory needs to be accessed frequently.
The SM utilization ranges between 25.26\% and 70.71\%, with a mean of 57.76\%.
The memory bandwidth utilization ranges between 19.63\% and 59.50\%, with a mean of 51.47\%.
Moreover, the SIMD utilization ranges between 66.4\% and 91.8\%, with a mean of 74.5\%.

\begin{table}[t]
    \centering
    \caption{Comparison with the GPU baseline}\label{tab:gpu-baseline}
    \resizebox{\columnwidth}{!}{
        \input{fig/04-evaluation/tab-gpu-baseline.tex}
    }
\end{table}

\subsection{Load Balance}\label{sec:eval-load-balance}

\begin{figure*}[t]
    \centering
    \includegraphics[width=\textwidth]{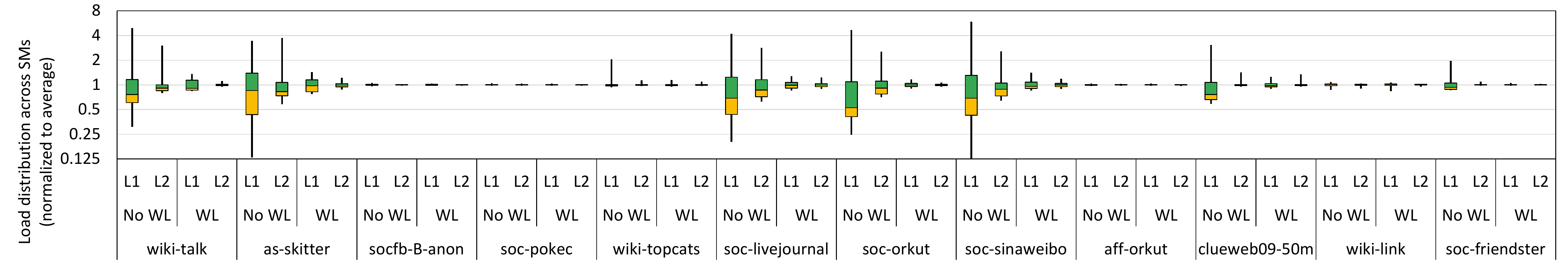}
    \caption{Load distribution across streaming multiprocessors (SMs) for different combinations of optimizations}
    \label{fig:mce-load-balance}
\end{figure*}

Fig.~\ref{fig:mce-load-balance} compares the distribution of load across SMs for the A100 GPU when different combinations of optimizations are applied.
The load of an SM is measured as the maximum number of tree nodes visited by any block on that SM (recall that we launch exactly the maximum number of concurrent blocks that can execute and reuse these blocks to process different subtrees).
Based on these results, we make three key observations.

The first observation is that when no worker list is used (No WL), using independent second-level subtrees (L2) instead of independent first-level subtrees (L1) substantially reduces load imbalance.
This observation is consistent with our prior work on $k$-clique counting~\cite{ourkclique}.
However, unlike our prior work, we note that in the case of MCE, even after L2 trees are used, the imbalance is still high for some graphs.
The average across benchmarks of the maximum load across thread blocks is 2.28$\times$ the average load when using L1 trees and 1.63$\times$ the average load when using L2 trees, which is a 1.40$\times$ decrease in imbalance.

The second observation is that using a worker list (WL) substantially reduces load imbalance compared to not using a worker list.
To further study the effectiveness of the worker list, Table~\ref{tab:donor-block} shows the number of donations performed for each graph.
It is clear that the graphs with a large number of donations are also the ones with high imbalance in Fig.~\ref{fig:mce-load-balance} that benefit from using the worker list.
These results show the effectiveness of our proposed worker list approach at reducing the load imbalance of MCE on GPUs.

The third observation is that when a worker list is used, there is little difference in load imbalance between using L1 trees and L2 trees in most cases.
The average across benchmarks of the maximum load across thread blocks is 1.17$\times$ the average load when using L1 trees and 1.11$\times$ the average load when using L2 trees, which is only a 1.05$\times$ decrease in imbalance.
This observation shows that our proposed worker list approach obviates the need to use L2 trees for the purpose of load balancing in most cases.
Still, using L2 trees may have other benefits such as smaller induced subgraphs and shorter set operations.
We revisit this point in Section~\ref{sec:eval-breakdown}.

\begin{table}[t]
    \centering
    \caption{Number of donations with a worker list}\label{tab:donor-block}
    \resizebox{\columnwidth}{!}{
        \input{fig/04-evaluation/donation-count}
    }
\end{table}

\begin{figure*}[t]
    \centering
    \includegraphics[width=\textwidth]{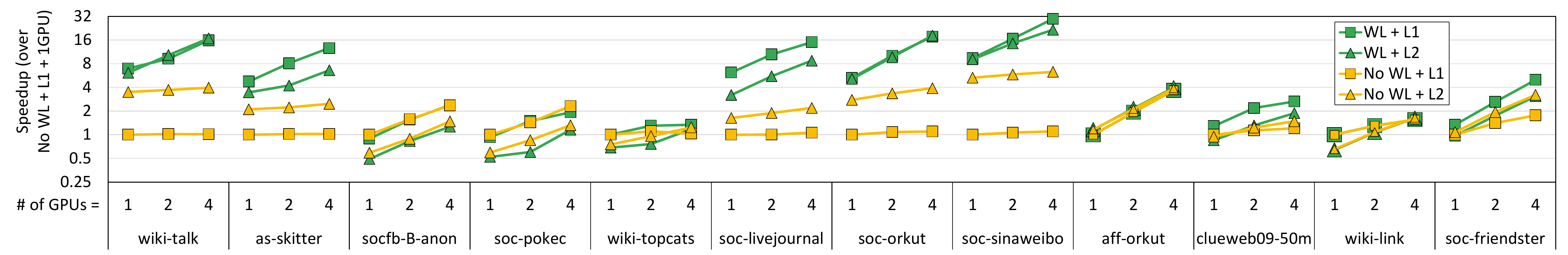}
    \caption{Strong scaling with respect to the number of GPUs for different combinations of optimizations}\label{fig:mce-multigpu}
\end{figure*}

\subsection{Scalability to Multiple GPUs}\label{sec:eval-scalability}

Fig.~\ref{fig:mce-multigpu} shows the strong scaling of our GPU implementation across multiple V100 GPUs when different combinations of optimizations are applied.
In the multi-GPU implementation, L1 or L2 trees are distributed across GPUs in a round-robin scheme and each GPU maintains its own private worker list.
We also experimented with using an inter-GPU shared worker list, however its overhead was not worth its benefit.
Currently, our implementation supports scaling to multiple GPUs within a single node, however, scaling to multiple GPU nodes is the subject of future work.

\begin{figure*}[t]
    \centering
    \includegraphics[width=\textwidth]{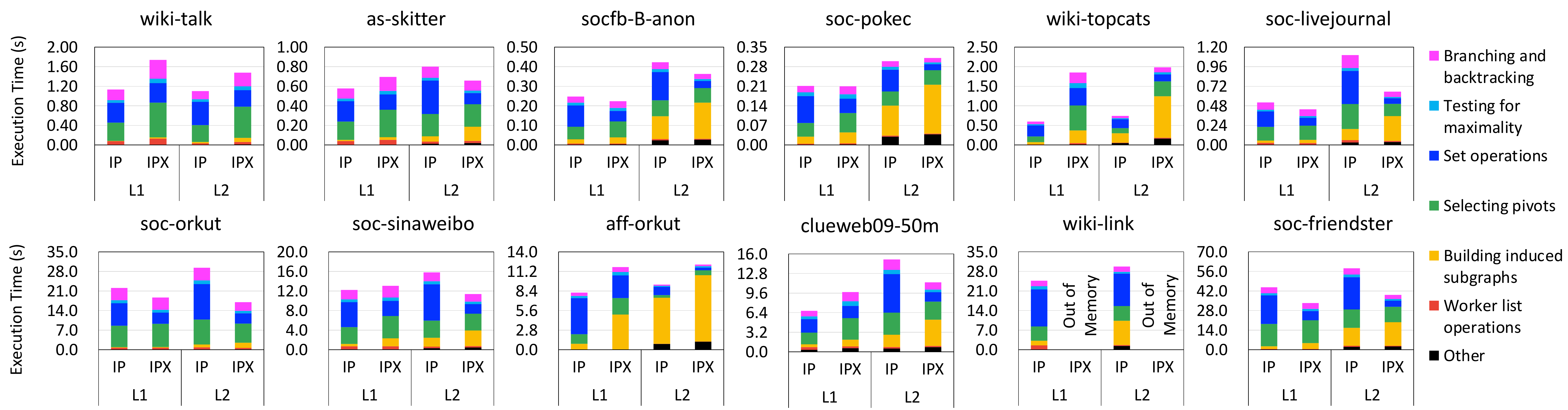}
    \caption{Breakdown and comparison of execution time for different combinations of optimizations}
    \label{fig:mce-breakdown}
\end{figure*}

Based on the results in Fig.~\ref{fig:mce-multigpu}, we make three key observations.
The first observation is that in most cases, on a single GPU, the implementations that use a worker list substantially outperform those that do not use a worker list.
This observation shows the effectiveness of our proposed worker list approach at improving performance by reducing load imbalance.

The second observation is that in most cases, as we scale to multiple GPUs, the WL implementation scales well whereas the No WL implementation scales poorly.
This observation shows that scaling to multiple GPUs exacerbates the load imbalance challenge of MCE, and that our proposed worker list approach is effective at overcoming this scalability challenge.

The third observation is that in most cases, using L2 trees instead of L1 trees has better performance and scalability when no worker list is used, but does not significantly improve performance and scalability when a worker list \textit{is} used and may even degrade performance.
This observation reiterates the observation in Section~\ref{sec:eval-load-balance} that our proposed worker list approach obviates the need to use L2 trees for the purpose of load balancing.

\subsection{Choice of Optimizations}\label{sec:eval-breakdown}

Fig.~\ref{fig:mce-breakdown} shows the breakdown of execution time on the A100 GPU when different combinations of optimizations are applied.
SM clocks are used to get the number of cycles spent by each thread block on each activity.

\textbf{L1 trees vs. L2 trees.}
When comparing the use of L1 trees with L2 trees, we make two key observations.
The first observation is that in most cases, the fraction of time spent on constructing the induced subgraph is larger for L2 trees.
The reason is that using L2 trees extracts more subtrees that are each smaller in size, so more induced subgraphs are generated and the cost of generating them is amortized across fewer tree node traversals.
The second observation is that in most cases, the fraction of time spent on set operations (such as intersections) is smaller for L2 trees.
The reason is that using L2 trees results in smaller induced subgraphs, hence smaller sets to operate on.
Nevertheless, the benefit of faster set operations does not overcome the increased overhead of constructing induced subgraphs, so we find that using L1 trees outperforms using L2 trees in the majority of cases.
On average, using L1 trees is 1.2$\times$ (geometric mean) faster than using L2 trees.
Note that in our prior work on $k$-clique counting~\cite{ourkclique}, L2 trees were more effective in most cases because of their load balancing benefits.
However, since these benefits are obviated by the worker list (see Sections~\ref{sec:eval-load-balance} and~\ref{sec:eval-scalability}), the benefits of L1 trees become more pronounced.

\textbf{Partial vs. full induced subgraphs.}
When comparing the use of partial induced subgraphs (IP, i.e., induced on $P$ only) and full induced subgraphs (IPX, i.e., induced on $P$ and $X$), we make three key observations.
The first observation is that in most cases, the fraction of time spent on constructing induced subgraphs is larger for IPX.
The reason is that the induced subgraphs in IPX are larger than the induced subgraphs in IP, thereby taking longer to construct.
The second observation is that in most cases, the fraction of time spent on pivoting is larger for IPX.
The reason is that in IPX, we consider pivots from $X \cup P$, whereas in IP, we only consider pivots in $X_P \cup P$ and do not consider pivots from $X_X$.
As a result, we spend less time on pivoting in IP, however, we may not find the best possible pivot.
The third observation is that in most cases, the fraction of time spent on set operations is smaller for IPX.
The reason is that including the edges between $P$ and $X$ vertices in the induced subgraphs makes set operations on the $X$ sets less costly.
This trade-off between the time spent on constructing induced subgraphs, the time spent on pivoting, and the time spent on performing set operations makes each approach perform better on different datasets.
Overall, IP performs better in six cases whereas IPX performs better in six cases.

\textbf{Heuristics for selecting optimizations.}
To select the best combination of optimizations, we recommend the following heuristic.
First, L1 trees should always be selected instead of L2 trees.
Second, IP should be selected when $\Delta/d > 200$, and IPX should be selected otherwise.
The intuition is that IPX requires $O(\Delta \cdot d)$ space for each induced subgraph and IP requires only $O(d^2)$ space, so $\Delta/d$ represents how much more space IPX requires compared to IP.
If this value is too high (IPX requires too much more space), it is better to select IP, otherwise it is better to select IPX.
Table~\ref{tab:ip-heuristic} shows that this heuristic selects the best combination in the majority of cases, with a geometric mean slowdown of 1.02$\times$ (up to 1.10$\times$) from selecting incorrectly.
One possible additional optimization is to use different strategies for storing different induced subgraphs within the same graph based on the local ratio of degree to out-degree.
This optimization is the subject of future work.

Finally, we make one additional observation from Fig.~\ref{fig:mce-breakdown} that the fraction of time spent adding to and removing from the worker list is small.
This observation shows that the substantial load balancing benefits that the worker list provides come with a low performance overhead.
Furthermore, the fraction of time spent performing worker list operations in Fig.~\ref{fig:mce-breakdown} tends to be larger for graphs where a large number of donations is performed according to Table~\ref{tab:donor-block}.

\useunder{\uline}{\ul}{}
\begin{table}[t]
\caption{Comparing heuristic (underlined) and optimal (bold) selection of optimization combinations}\label{tab:ip-heuristic}
\resizebox{\columnwidth}{!}{
\begin{tabular}{lrrrrr|r|}
\cline{3-7}
 & \multicolumn{1}{l|}{} & \multicolumn{4}{c|}{Execution time (s)} & \multicolumn{1}{c|}{\multirow{2}{*}{\begin{tabular}[c]{@{}c@{}}Heuristic \\ slowdown\end{tabular}}} \\ \cline{1-6}
\multicolumn{1}{|c|}{Graph} & \multicolumn{1}{c|}{$\Delta$/ $d$} & \multicolumn{1}{c|}{L1 + IP} & \multicolumn{1}{c|}{L1 + IPX} & \multicolumn{1}{c|}{L2 + IP} & \multicolumn{1}{c|}{L2 + IPX} & \multicolumn{1}{c|}{} \\ \hline
\multicolumn{1}{|l|}{wiki-talk} & \multicolumn{1}{r|}{763.58} & \multicolumn{1}{r|}{{\ul 1.14}} & \multicolumn{1}{r|}{1.74} & \multicolumn{1}{r|}{\textbf{1.10}} & 1.48 & 1.03 \\ \hline
\multicolumn{1}{|l|}{as-skitter} & \multicolumn{1}{r|}{319.41} & \multicolumn{1}{r|}{{\ul \textbf{0.58}}} & \multicolumn{1}{r|}{0.70} & \multicolumn{1}{r|}{0.80} & 0.66 & 1.00 \\ \hline
\multicolumn{1}{|l|}{socfb-B-anon} & \multicolumn{1}{r|}{69.14} & \multicolumn{1}{r|}{{0.25}} & \multicolumn{1}{r|}{{\ul \textbf{0.22}}} & \multicolumn{1}{r|}{0.42} & 0.36 & 1.00 \\ \hline
\multicolumn{1}{|l|}{soc-pokec} & \multicolumn{1}{r|}{316.04} & \multicolumn{1}{r|}{{\ul 0.21}} & \multicolumn{1}{r|}{\textbf{0.21}} & \multicolumn{1}{r|}{0.30} & 0.31 & 1.00 \\ \hline
\multicolumn{1}{|l|}{wiki-topcats} & \multicolumn{1}{r|}{2,407.49} & \multicolumn{1}{r|}{{\ul \textbf{0.60}}} & \multicolumn{1}{r|}{1.85} & \multicolumn{1}{r|}{0.74} & 1.99 & 1.00 \\ \hline
\multicolumn{1}{|l|}{soc-livejournal} & \multicolumn{1}{r|}{12.45} & \multicolumn{1}{r|}{{0.52}} & \multicolumn{1}{r|}{{\ul \textbf{0.44}}} & \multicolumn{1}{r|}{1.10} & 0.65 & 1.00 \\ \hline
\multicolumn{1}{|l|}{soc-orkut} & \multicolumn{1}{r|}{131.67} & \multicolumn{1}{r|}{{22.10}} & \multicolumn{1}{r|}{{\ul 18.68}} & \multicolumn{1}{r|}{29.29} & \textbf{16.98} & 1.10 \\ \hline
\multicolumn{1}{|l|}{soc-sinaweibo} & \multicolumn{1}{r|}{1,442.95} & \multicolumn{1}{r|}{{\ul 12.20}} & \multicolumn{1}{r|}{13.05} & \multicolumn{1}{r|}{15.80} & \textbf{11.39} & 1.07 \\ \hline
\multicolumn{1}{|l|}{aff-orkut} & \multicolumn{1}{r|}{675.73} & \multicolumn{1}{r|}{{\ul \textbf{8.16}}} & \multicolumn{1}{r|}{11.82} & \multicolumn{1}{r|}{9.31} & 12.18 & 1.00 \\ \hline
\multicolumn{1}{|l|}{clueweb09-50m} & \multicolumn{1}{r|}{1,606.65} & \multicolumn{1}{r|}{{\ul \textbf{6.70}}} & \multicolumn{1}{r|}{9.77} & \multicolumn{1}{r|}{15.10} & 11.35 & 1.00 \\ \hline
\multicolumn{1}{|l|}{wiki-link} & \multicolumn{1}{r|}{3,813.70} & \multicolumn{1}{r|}{{\ul \textbf{24.67}}} &\multicolumn{1}{c|}{\begin{tabular}[c]{@{}l@{}}-\end{tabular}} & \multicolumn{1}{r|}{29.76} &\multicolumn{1}{c|}{\begin{tabular}[c]{@{}l@{}}-\end{tabular}} & 1.00 \\ \hline
\multicolumn{1}{|l|}{soc-friendster} & \multicolumn{1}{r|}{17.15} & \multicolumn{1}{r|}{{44.64}} & \multicolumn{1}{r|}{{\ul \textbf{33.41}}} & \multicolumn{1}{r|}{58.16} & 39.17 & 1.00 \\ \hline
 & \multicolumn{1}{l}{} & \multicolumn{1}{l}{} & \multicolumn{1}{l}{} & \multicolumn{1}{l|}{} & \multicolumn{1}{l|}{Geomean} & 1.02 \\ \cline{6-7} 
\end{tabular}
}
\end{table}

%% file: fig/04-evaluation/tab-dataset.tex
\begin{tabular}{|l|r|r|r|r|r|r|r|rr|rr|rr|}
\hline
 & \multicolumn{1}{c|}{} & \multicolumn{1}{c|}{} & \multicolumn{1}{c|}{} & \multicolumn{1}{c|}{} & \multicolumn{1}{c|}{} & \multicolumn{1}{c|}{{\color[HTML]{000000} }} & \multicolumn{1}{c|}{{\color[HTML]{000000} }} & \multicolumn{2}{c|}{{\color[HTML]{000000} Parallel CPU time (s)}} & \multicolumn{2}{c|}{\begin{tabular}[c]{@{}c@{}}GPU \\ time (s)\end{tabular}} & \multicolumn{2}{c|}{\begin{tabular}[c]{@{}c@{}}GPU speedup over \\ Skylake with 96 threads \end{tabular}} \\ \cline{9-14} 
\multirow{-2}{*}{Graph} & \multicolumn{1}{c|}{\multirow{-2}{*}{$|V|$}} & \multicolumn{1}{c|}{\multirow{-2}{*}{$|E|$}} & \multicolumn{1}{c|}{\multirow{-2}{*}{\begin{tabular}[c]{@{}c@{}}Max degree \\ ($\Delta$)\end{tabular}}} & \multicolumn{1}{c|}{\multirow{-2}{*}{\begin{tabular}[c]{@{}c@{}}Degeneracy \\ ($d$)\end{tabular}}} & \multicolumn{1}{c|}{\multirow{-2}{*}{\begin{tabular}[c]{@{}c@{}}\# of maximal \\ cliques\end{tabular}}} & \multicolumn{1}{c|}{\multirow{-2}{*}{{\color[HTML]{000000} \begin{tabular}[c]{@{}c@{}}Avg maximal \\ clique size\end{tabular}}}} & \multicolumn{1}{c|}{\multirow{-2}{*}{{\color[HTML]{000000} \begin{tabular}[c]{@{}c@{}}Max \\ clique size\end{tabular}}}} & \multicolumn{1}{c|}{{\color[HTML]{000000} \begin{tabular}[c]{@{}l@{}}Cascade Lake \\ with 80 threads\end{tabular}}} & \multicolumn{1}{c|}{\begin{tabular}[c]{@{}c@{}}Skylake with \\ 96 threads~\cite{mcc6_baseline}\end{tabular}} & \multicolumn{1}{c|}{V100} & \multicolumn{1}{c|}{A100} & \multicolumn{1}{c|}{\quad \hspace{0.3em} V100 \hspace{0.3em} \quad} & \multicolumn{1}{c|}{A100} \\ \hline
wiki-talk~\cite{snap} & 2,394,385 & 4,659,565 & 100,029 & 131 & 86,333,306 & {\color[HTML]{000000} 13.37} & {\color[HTML]{000000} 26} & \multicolumn{1}{r|}{{\color[HTML]{000000} 4.57}} & 4 & \multicolumn{1}{r|}{1.39} & 1.38 & \multicolumn{1}{r|}{2.89} & 2.91 \\ \hline
as-skitter~\cite{snap} & 1,696,415 & 11,095,298 & 35,455 & 111 & 37,322,355 & {\color[HTML]{000000} 19.91} & {\color[HTML]{000000} 67} & \multicolumn{1}{r|}{{\color[HTML]{000000} 3.74}} & 3 & \multicolumn{1}{r|}{0.81} & 0.79 & \multicolumn{1}{r|}{3.70} & 3.82 \\ \hline
socfb-B-anon~\cite{netrepo} & 2,937,613 & 20,959,854 & 4,356 & 63 & 27,593,398 & {\color[HTML]{000000} 5.24} & {\color[HTML]{000000} 24} & \multicolumn{1}{r|}{{\color[HTML]{000000} 2.38}} & 2 & \multicolumn{1}{r|}{0.56} & 0.45 & \multicolumn{1}{r|}{3.60} & 4.42 \\ \hline
soc-pokec~\cite{snap} & 1,632,804 & 22,301,964 & 14,854 & 47 & 19,376,873 & {\color[HTML]{000000} 3.67} & {\color[HTML]{000000} 29} & \multicolumn{1}{r|}{{\color[HTML]{000000} 1.45}} & 1 & \multicolumn{1}{r|}{0.38} & 0.36 & \multicolumn{1}{r|}{2.65} & 2.78 \\ \hline
wiki-topcats~\cite{snap} & 1,791,489 & 25,444,207 & 238,342 & 99 & 27,229,873 & {\color[HTML]{000000} 4.46} & {\color[HTML]{000000} 39} & \multicolumn{1}{r|}{{\color[HTML]{000000} 2.09}} & 2 & \multicolumn{1}{r|}{0.83} & 0.87 & \multicolumn{1}{r|}{2.42} & 2.30 \\ \hline
soc-livejournal~\cite{netrepo} & 4,033,138 & 27,933,062 & 2,651 & 213 & 38,413,665 & {\color[HTML]{000000} 29.97} & {\color[HTML]{000000} 214} & \multicolumn{1}{r|}{{\color[HTML]{000000} 5.45}} & 5 & \multicolumn{1}{r|}{0.81} & 0.76  & \multicolumn{1}{r|}{6.21} & 6.57 \\ \hline
soc-orkut~\cite{netrepo} & 3,072,442 & 117,185,083 & 33,313 & 253 & 2,269,631,973 & {\color[HTML]{000000} 20.24} & {\color[HTML]{000000} 51} & \multicolumn{1}{r|}{{\color[HTML]{000000} 110.61}} & 93 & \multicolumn{1}{r|}{25.23} & 17.82 & \multicolumn{1}{r|}{3.69} & 5.22 \\ \hline
soc-sinaweibo~\cite{netrepo} & 58,655,850 & 261,321,033 & 278,489 & 193 & 1,117,416,174 & {\color[HTML]{000000} 18.43} & {\color[HTML]{000000} 44} & \multicolumn{1}{r|}{{\color[HTML]{000000} 67.78}} & 54 & \multicolumn{1}{r|}{16.40} & 13.60 & \multicolumn{1}{r|}{3.29} & 3.97 \\ \hline
aff-orkut~\cite{netrepo} & 8,730,858 & 327,036,486 & 318,268 & 471 & 417,032,363 & {\color[HTML]{000000} 2.53} & {\color[HTML]{000000} 6} & \multicolumn{1}{r|}{{\color[HTML]{000000} 138.89}} & 147 & \multicolumn{1}{r|}{14.40} & 8.82 & \multicolumn{1}{r|}{10.21} & 16.67 \\ \hline
clueweb09-50m~\cite{netrepo} & 428,136,613 & 446,766,953 & 308,477 & 192 & 1,001,323,679 & {\color[HTML]{000000} 15.21} & {\color[HTML]{000000} 56} & \multicolumn{1}{r|}{{\color[HTML]{000000} 99.29}} & 90 & \multicolumn{1}{r|}{14.90} & 10.03 & \multicolumn{1}{r|}{6.04} & 8.97 \\ \hline
wiki-link~\cite{netrepo} & 27,154,799 & 543,183,611 & 4,271,341 & 1,120 & 568,730,123 & {\color[HTML]{000000} 4.51} & {\color[HTML]{000000} 428} & \multicolumn{1}{r|}{{\color[HTML]{000000} 112.14}} & 109 & \multicolumn{1}{r|}{34.82} & 32.23 & \multicolumn{1}{r|}{3.13} & 3.38 \\ \hline
soc-friendster~\cite{netrepo} & 65,608,367 & 1,806,067,135 & 5,214 & 304 & 3,364,773,700 & {\color[HTML]{000000} 6.88} & {\color[HTML]{000000} 129} & \multicolumn{1}{r|}{{\color[HTML]{000000} 406.33}} & 380 & \multicolumn{1}{r|}{64.50} & 39.59 & \multicolumn{1}{r|}{5.89} & 9.60 \\ \hline
\end{tabular}%

%% file: fig/04-evaluation/tab-gpu-baseline.tex
\begin{tabular}{|l|r|r|r|}
\hline
    & GPU baseline on & \multicolumn{1}{c|}{Our implementation} & \multicolumn{1}{c|}{Speedup over} \\
    \multirow{-2}{*}{Graph} & \multicolumn{1}{c|}{Titan X (s) \cite{mcg6_2021}} & \multicolumn{1}{c|}{on V100 (s)} & \multicolumn{1}{c|}{GPU baseline} \\
    \hline wiki-talk & 41.09 & 1.39 & 29.56 \\
    \hline as-skitter & 40.87 & 0.81 & 50.46 \\
    \hline soc-pokec & 12.85 & 0.38 & 33.82 \\
    \hline wiki-topcats & 26.57 & 0.83 & 32.01 \\
    \hline
\end{tabular}

%% file: fig/04-evaluation/donation-count.tex
\begin{tabular}{|l|r|r|l|l|r|r|}
\cline{1-3}\cline{5-7}
    Graph & \multicolumn{1}{c|}{L1} & \multicolumn{1}{c|}{L2} &   & Graph & \multicolumn{1}{c|}{L1} & \multicolumn{1}{c|}{L2}  \\ \cline{1-3}\cline{5-7}
    wiki-talk & 1,436,113 & 74,268 &   & soc-orkut & 9,728,983 & 2,438,166  \\ \cline{1-3}\cline{5-7}
    as-skitter & 625,508 & 74,703 &   & soc-sinaweibo & 11,119,926 & 1,164,829  \\ \cline{1-3}\cline{5-7}
    socfb-b-anon & 170 & 0 &   & aff-orkut & 0 & 0  \\ \cline{1-3}\cline{5-7}
    soc-pokec & 0 & 0 &   & clueweb09-50m & 890,122 & 49,799  \\ \cline{1-3}\cline{5-7}
    wiki-topcats & 2,128 & 28 &   & wiki-link & 2,997 & 99  \\ \cline{1-3}\cline{5-7}
    soc-livejournal & 341,721 & 105,270 &   & soc-friendster & 1,615,437 & 358,499  \\ \cline{1-3}\cline{5-7}
\end{tabular}

%% file: sec/5-related.tex
\section{Related Work}

MCE has been extensively studied on CPUs~\cite{bk1973, bk_tomita, bk_eppstein,mcg4_prev, mcc7} and many attempts to parallelize it on the CPU have been made~\cite{mcg5, mcg4, mcc6_baseline, mcc2, mcc3, mcc1}.
To the best of our knowledge, the work of Blanu{\v{s}}a et al.~\cite{mcc6_baseline} is the state-of-the-art parallel CPU implementation of MCE, and its reported performance is the highest among all prior CPU (and GPU) implementations.
Our work targets accelerating MCE on GPUs.
We compare the performance of our work to that of Blanu{\v{s}}a et al.~\cite{mcc6_baseline} in Section~\ref{sec:eval}.
MCE has also been parallelized on distributed CPU systems~\cite{mcc4, distMCE1, distMCE2, distMCE3}.
Our work focuses on parallelizing MCE on single-node single- and multi-GPU systems, however, parallelizing MCE on distributed GPU systems is an interesting future work.
Many works have parallelized MCE on GPUs~\cite{mcg1, lessons, mcg3, mcg2, mcg6_2021}.
We compare our approach to these works in depth in Section~\ref{sec:challenges}.

$k$-clique enumeration has been studied on CPUs~\cite{chiba1985arboricity, degreeOrdering1, kclist, arb-count, orderingheuristic, gianinazzi2021parallel, lonkar2021accelerating} and GPUs~\cite{ourkclique}.
Triangle counting, which is a special case of $k$-clique counting, has also been studied on CPUs~\cite{tccpu1, tccpu2, tccpu3, tccpu4, steil2021tripoll} and GPUs~\cite{almasri2021hykernel, tc1, tc2, tc3, tc4, tc5, tc6, tc7, tc8, tc9,olabi2022compiler}.
Our MCE work uses similar techniques to those used in our prior GPU work on $k$-clique counting~\cite{ourkclique}, namely per-block depth-first traversal, binary encoding of induced subgraphs, and subwarp partitioning.
However, as discussed in Section~\ref{sec:challenges}, MCE imposes unique challenges that we overcome with additional techniques such as the worker list, partial induced subgraphs, and the compact representation of excluded vertex sets.

Generalized graph pattern matching has also been studied on CPUs~\cite{escape, motif1, motif2, motif3, motif4} and GPUs~\cite{pangolin, sub0, sub1, sub2, sub3, sub4, sub5, chen2021efficient, pbe}.
Cliques are special cases of patterns that graph pattern matching works aim to find.
While graph pattern matching algorithms perform similar tree traversals to those in MCE, their general nature makes them more difficult to scale.
For example, using induced subgraphs in graph pattern matching would require $O(\Delta^2)$ space, which causes most of these works to avoid such an optimization.

$k$-truss decomposition has also been studied on CPUs~\cite{cputruss0, cputruss1, cputruss2, cputruss3} and GPUs~\cite{truss0, truss1, truss2, truss3, truss4, truss5}.
A truss is a relaxation of a clique, and finding trusses uses significantly different techniques that are not based on search trees.
In our work, we aim to find exact maximal cliques.

%% file: sec/6-conculsion.tex
\section{Conclusion}

We present a GPU solution for accelerating maximal clique enumeration that assigns independent subtrees to different thread blocks and has each thread block perform a depth-first traversal of its subtree.
We propose a worker list for dynamic load balancing to mitigate the high imbalance in the MCE search tree.
We propose partial induced subgraphs and a compact representation of excluded vertex sets to regulate memory consumption.
We also apply various optimizations used in prior works such as binary encoding of induced subgraph and partitioning work at subwarp granularity.
Our evaluation shows that our GPU implementation substantially outperforms the state-of-the-art parallel CPU implementation, which outperforms prior GPU implementations.

%% file: sec/ack.tex
\section*{Acknowledgment}

We thank Zaid Qureshi and Amir Nassereldine for their insights and technical assistance.
This work is supported in part by the IBM-Illinois Center for Cognitive Computing Systems Research (C3SR).
Izzat El Hajj acknowledges the support of the University Research Board of the American University of Beirut (AUB-URB-104391-26749). Jinjun Xiong acknowledges the support of NSF (FuSe-TG 2235364) and the joint support of NSF and IES through AI4ExceptionalEd (2229873).
We are also grateful to NVIDIA's Applied Research Accelerator Program for donating A100 GPUs that were helpful in the final testing stages of this work.

%% file: sec/artifact.tex
\appendix[Artifact Appendix]

\subsection{Abstract}

The artifact contains a pre-compiled binary for our application, a Dockerfile preparing software dependencies, and scripts for downloading datasets, running experiments, and reproducing figures and tables in Section~\ref{sec:eval}. To reproduce results inside the docker image built from the Dockerfile, a CPU with 4 cores in x86\_64 architecture, 128 GB of RAM, 256 GB of disk space
and 4 NVIDIA GPUs of compute capability 7.0 or higher (i.e., Volta architecture or later) with 32 GB of GPU memory each are the minimum hardware requirements. We also require CUDA driver version of at least 450.80.02 with built-in CUB library, or CUDA driver version of at least 440.33 with CUB library from source in Linux OS, and Docker version above 19.03 with NVIDIA Container Toolkit as the run-time environment. Our source code is also provided in case there is a version mismatch in the environment, which may require recompilation.

\subsection{Artifact check-list (meta-information)}

{\small
\begin{itemize}
  \item {\bf Algorithm:} Bron-Kerbosch Algorithm for maximal clique enumeration on GPUs
  \item {\bf Binary:} A pre-compiled binary built from the Makefile is provided, with software dependencies prepared in the Dockerfile.
  \item {\bf Dataset:} The SNAP Datasets\footnote{\href{https://snap.stanford.edu/}{https://snap.stanford.edu/}} and the Network Repository\footnote{\href{https://networkrepository.com/}{https://networkrepository.com/}}
  \item {\bf Run-time environment:} CUDA driver version of at least 450.80.02 with built-in CUB library, or CUDA driver version of at least 440.33 with CUB library from source\footnote{\href{https://github.com/NVIDIA/cub}{https://github.com/NVIDIA/cub}} in Linux OS, and Docker version above 19.03 with NVIDIA Container Toolkit\footnote{\href{https://docs.nvidia.com/datacenter/cloud-native/container-toolkit/latest/}{https://docs.nvidia.com/datacenter/cloud-native/container-toolkit/latest/}}
  \item {\bf Hardware:} At least a CPU with 4 cores in x86\_64 architecture, 128 GB of RAM, 256 GB of disk space
and 4 NVIDIA GPUs of compute capability 7.0 or higher (i.e., Volta architecture or later) with 32 GB of GPU memory each
  \item {\bf Execution:} Approximately 6 hours, might fluctuate based on downloading datasets in the first hour due to internet bandwidth
  \item {\bf Output:} Figures and tables in Section~\ref{sec:eval}
  \item {\bf Experiments:} Building docker image, downloading datasets, running experiments, and visualizing results as figures and tables
  \item {\bf Disk space required:} 200 GB
  \item {\bf Publicly available: } \href{https://github.com/yen-hsiang-chang/mce-gpu}{https://github.com/yen-hsiang-chang/mce-gpu}
  \item {\bf Code license:} University of Illinois/NCSA Open Source License
  \item {\bf Archived:} \href{https://zenodo.org/record/8270171}{https://zenodo.org/record/8270171}
\end{itemize}
}

\subsection{Description}

\subsubsection{How to access}

The artifact can be accessed from GitHub at \href{https://github.com/yen-hsiang-chang/mce-gpu}{https://github.com/yen-hsiang-chang/mce-gpu}.

\subsubsection{Hardware dependencies}

Our experiments require at least a CPU with 4 cores in x86\_64 architecture to have a unique CPU thread for each GPUs, 128 GB of RAM to load and unzip graphs, 256 GB of disk space to store datasets,
and 4 NVIDIA GPUs of compute capability 7.0 or higher (i.e., Volta architecture or later) with 32 GB of GPU memory each to execute our kernels on multiple GPUs. Our multi-GPU experiments are done on a platform with four 32GB NVIDIA V100 GPUs attached to an Intel Xeon Gold 6230 CPU. Other platform satisfying the requirements will achieve similar results, while the performance might fluctuate due to different computing power and memory bandwidth.

\subsubsection{Software dependencies}

We require CUDA driver version of at least 450.80.02 with built-in CUB library, or CUDA driver version of at least 440.33 with CUB library from source to pre-process the graphs. We also require Docker 19.03 or higher with NVIDIA Container Toolkit to make GPUs ready to be used with Docker. Other software dependencies prepared in the docker image include OpenMP 4.5 to have a unique CPU thread for each GPUs, and Python 3.6 with numpy, matplotlib and tabulate to run the pre-compiled binary, plotting figures and formatting tables.

\subsubsection{Datasets}

We use wiki-talk, as-skitter, soc-pokec and wiki-topcats from the SNAP datasets and socfb-B-anon, soc-livejournal, soc-orkut, soc-sinaweibo, aff-orkut, clueweb09-50m, wiki-link and soc-friendster from the Network Repository as our datasets for evaluation. A script is provided in the artifact to help download and extract datasets.

\subsection{Installation and Experiment Instructions}

\begin{enumerate}
    \item Install Docker with NVIDIA Container Toolkit and install CUDA driver with CUB library
    \item Get the artifact from GitHub
    \item Build and launch the docker image: \\ \texttt{\$ ./docker.sh /path/to/data/ \symbol{92} \\ /path/to/results/} \\
    Notice that \texttt{/path/to/data/} is the path on host that will store datasets and it needs to have at least 200 GB disk space, and \texttt{/path/to/results/} is the path on host that will store evaluation results
    \item Inside the docker image, reproduce all experiments with the script provided: \\ \texttt{\$ ./all\_experiments.sh} \\
    The whole experiments take about six hours to finish, and there might be some fluctuations since downloading datasets depends on the internet bandwidth. The \texttt{all\_experiments.sh} script includes downloading datasets using \texttt{download.py}, running experiments and evaluations in load balance, time breakdown and donation using \texttt{run.py}, and plotting figures and formatting tables using \texttt{plot.py}
    \item After the experiments are done, exit the docker image and inspect the results in \texttt{/path/to/results/} on host
\end{enumerate}

\subsection{Evaluation and expected results}
The \texttt{/path/to/results/} directory on host contains figures and tables reported in Section~\ref{sec:eval}, where figures are stored in the \texttt{plot/} sub-directory and tables are stored in the \texttt{table/} sub-directory. The descriptions are as follows and please refer to the paper for more details:
\begin{enumerate}
    \item \texttt{load-balance.png}: Visualize distribution of loads across streaming multiprocessors (SMs) for different combinations of optimizations as in Fig.~\ref{fig:mce-load-balance}
    \item \texttt{multigpu.png}: Visualize strong scaling experiments for different combinations of optimizations as in Fig.~\ref{fig:mce-multigpu}
    \item \texttt{breakdown.png}: Visualize time breakdown of execution time for different combinations of optimizations as in Fig.~\ref{fig:mce-breakdown}
    \item \texttt{time.txt}: Output the GPU time as in Table~\ref{tab:dataset}. Note that the time includes both the degeneracy ordering time and the maximal clique counting time
    \item \texttt{donation.txt}: Output the number of donations as in Table~\ref{tab:donor-block}
    \item \texttt{heuristics.txt}: Output the GPU time for different combinations of optimizations as in Table~\ref{tab:ip-heuristic}

\end{enumerate}
We expect the results to be similar as our evaluation in Section~\ref{sec:eval}. However, we do expect some minor differences for the GPU time reported in Fig.~\ref{fig:mce-breakdown} and Table~\ref{tab:ip-heuristic} in the paper, as optimization combinations are sensitive to memory bandwidth and computing power on GPUs, and different GPUs have different characteristics.

\subsection{Notes}

Our code has been open-sourced on GitHub to enable further research on accelerating maximal clique enumeration on GPUs. The repository contains a README file with instructions on running experiments without Docker and usages of the pre-compiled binary and each scripts if running individually.